\documentclass[aps,prl,twocolumn,groupedaddress,showpacs,letterpaper,floatfix,10pt]{revtex4-1}
\usepackage{amsmath,amssymb,graphicx,hyperref}
\hypersetup{pdfnewwindow=true, colorlinks=true, linkcolor=blue, anchorcolor=blue,
  citecolor=blue, filecolor=blue, menucolor=blue, urlcolor=blue}

\def\avg#1{\langle #1 \rangle}

\def\xc{x_\text{c}}

\begin{document}

\title{Phase separation in doped Mott insulators}

\author{Chuck-Hou Yee}
\email{cyee@kitp.ucsb.edu}
\affiliation{Kavli Institute for Theoretical Physics, University of California Santa Barbara, CA 93106, USA}
\author{Leon Balents}
\affiliation{Kavli Institute for Theoretical Physics, University of California Santa Barbara, CA 93106, USA}

\date{\today}

\begin{abstract}
  Motivated by the commonplace observation of Mott insulators away from integer
  filling, we construct a simple thermodynamic argument for phase separation in
  first-order doping-driven Mott transitions. We show how to compute the
  critical dopings required to drive the Mott transition using electronic
  structure calculations for the titanate family of perovskites, finding good
  agreement with experiment. The theory predicts the transition is percolative
  and should exhibit Coulomb frustration.
\end{abstract}

\pacs{}

\maketitle

The Mott transition is a pervasive and complex phenomena, observed in many
correlated oxide systems~\cite{Imada1998}.  It comes in two varieties: the
bandwidth-controlled transition at half-filling, tuned by the ratio of the
on-site Coulomb repulsion $U$ and band-width $W$, and the filling-controlled
transition, tuned by electron doping $x$ away from half-filling.
Theoretically, Mott insulators exist only at half-filling: with one electron
per site, hoppings necessarily create empty and doubly-occupied sites which are
heavily penalized by $U$. Introducing a finite charge density allows carriers
to move without incurring the on-site Coulomb cost, destroying the Mott
insulator~\cite{Fazekas1999}. However, experiments in a wide variety of
transition metal oxides show that the critical doping $\xc$ needed to destroy
insulating transport is not zero, but rather a substatial fraction of
unity~\cite{Fujimori1992b}, ranging from $0.1$ in the
nickelates~\cite{Xiang2010} to $0.5$ in the vanadates~\cite{Belik2007}.
Systematic variations of $\xc$ with bandwidth also argue that it is an
intrinsic quantity\cite{Katsufuji1997}, and motivate the search for mechanisms
independent of disorder or coupling to lattice vibrations for insulating
behavior away from half-filling.

The scenario of doping a Mott insulator has been heavily studied by a variety
of techniques~\cite{Kajueter1996, Lee2006, Kotliar1986, Onoda2001}. For the
classic case of a square lattice, basic issues such as whether the Mott
transition is first~\cite{Balzer2009, Misawa2013a} or second~\cite{Georges1996,
  Werner2007, Furukawa1992} order, the specific parameter regimes and
underlying mechanisms of phase separation~\cite{Visscher1974, Emery1990,
  Putikka2000, White2000, Galanakis2011}, and the structure of the
inhomogeneous phases~\cite{Spivak2006, Ortix2009, Giuliani2011} have been
actively researched, with results dependant on the precise model considered. We
take a different approach: we assume the {\em bandwidth}-controlled Mott
transition is first-order and deduce its implications by constructing a simple
thermodynamic description. We predict that the filling-controlled transition is
first order as a consequence, implying that phase separation occurs and the
critical doping scales as $\xc \sim \sqrt{U-U_c}$, where $U_c$ defines the
critical $U$ for the bandwidth controlled transition. We show how to compute
$\xc$ in electronic structure calculations~\cite{Kotliar2006}, using the rare
earth titanates~\cite{Greedan1985, Mochizuki2004} as a protypical example.


\emph{Thermodynamics} -- We construct a theory of the Mott transition by
connecting the bandwidth- and filling-controlled transitions.  By assuming the
former transition is first-order (which covers the majority of cases observed
in experiment), we can explicitly write down the energy densities $\epsilon =
E/V$ for the metallic and insulating states, since the two states must
independently exist over a finite parameter range and cross at the first-order
transition. We determine the phase boundary of the Mott transition in the
$\mu$-$U$ plane ($\mu$ is chemical potential) and compute the scaling of the
critical doping $\xc$ with $U$.

\begin{figure}
  \includegraphics[width=\columnwidth]{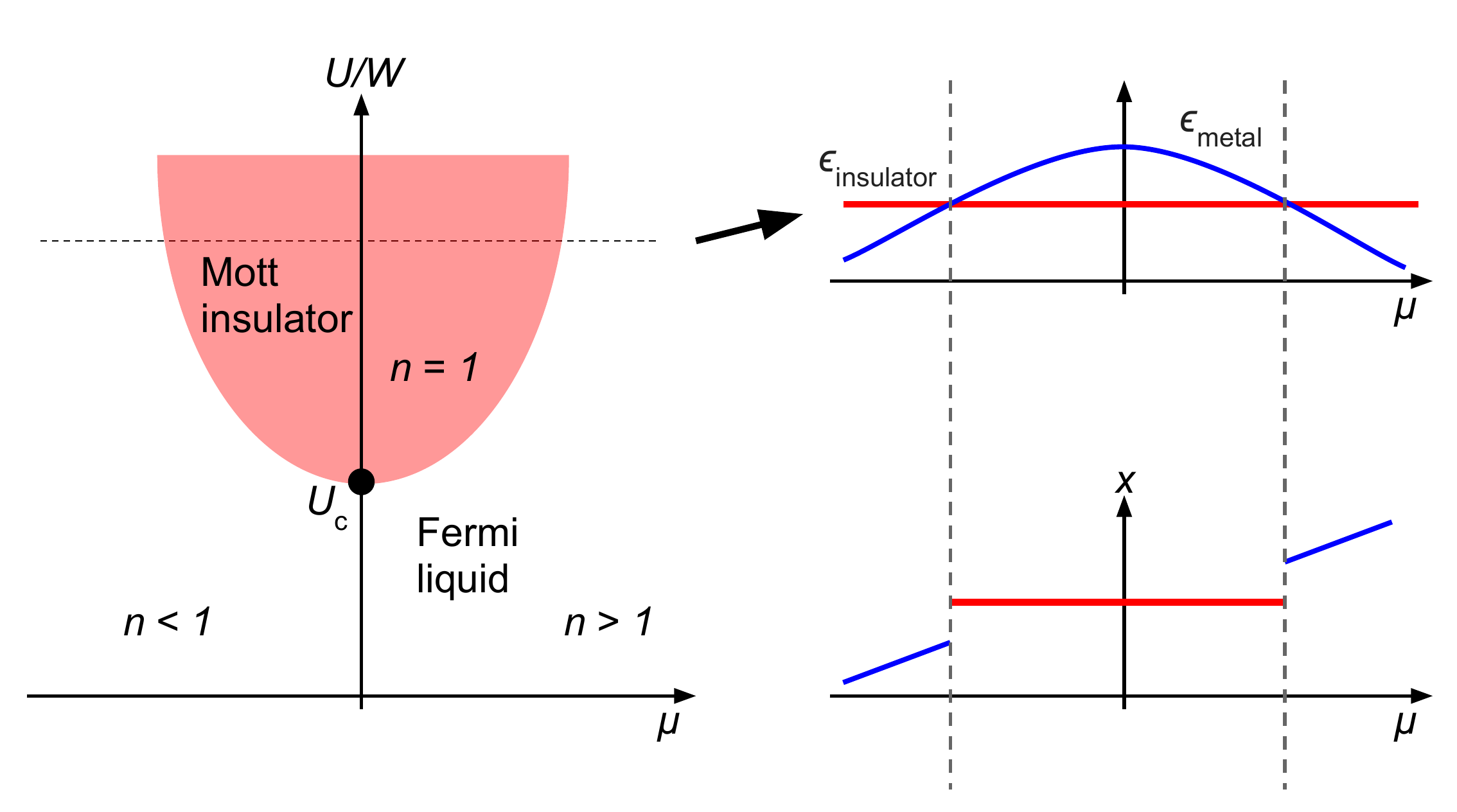}
  \caption{The phase diagram for the Mott transition, plotted as a function of
    interaction strength $U$ and chemical potential $\mu$. The energy vs. $\mu$
    curve at constant $U$ exhibits level crossings between the metallic and
    insulating states. The discontinuity in the derivative $x = -\partial
    \epsilon/\partial \mu$ implies thermodynamically forbidden densities where
    the system will phase separate into undoped $x = 0$ and critically-doped $x
    = \xc$ patches.}
  \label{fig:thermo}
\end{figure}

Consider a one-band Hubbard model on generic lattice. The $\mu$-$U$ phase
diagram generically consists of two regions: a Mott insulator occupying a
finite range in $\mu$ at sufficiently large $U > U_\text{c}$, and a Fermi
liquid (actually a superconductor or any other compressible phase including a
possible non-Fermi liquid will suffice for the argument) everywhere else
(Fig.~\ref{fig:thermo}). Expanding the grand-canonical energy densities of the
metal and insulator to lowest order about the bandwidth-controlled transition
point (dot in labeled $U_\text{c}$ in Fig.~\ref{fig:thermo}), we obtain:
\begin{align}
  \epsilon_\text{m}(\mu,U) &= \epsilon_0 + d_\text{m} \Delta U - \frac{1}{2} \kappa (\Delta\mu)^2 \\
  \epsilon_\text{i}(\mu,U) &= \epsilon_0 + d_\text{i} \Delta U.
\end{align}
Here, $\kappa = \partial x/\partial \mu$ is the electronic compressibility,
where the doping $x = n-1$ is defined relative to half-filling, and
$d_\text{m}$ and $d_\text{i}$ are the per-site double-occupancies
$\avg{n_{i\uparrow}n_{i\downarrow}}$ in the metallic and insulating states. The
chemical potential $\Delta\mu = \mu - \mu_{n=1}$ and Coulomb repulsion $\Delta U
= U - U_\text{c}$ are measured relative to the bandwidth-controlled transition
point.

Equating the two energies, we obtain the Mott phase boundary,
\begin{equation}
  \Delta U = \frac{\Delta\mu^2}{2} \frac{\kappa}{d_\text{m} - d_\text{i}}.\label{eq:1}
\end{equation}
The quadratic dependence $U \sim \mu^2$ is observed within
DMFT~\cite{Werner2007}. Evaluating the metallic density $x = -\partial
\epsilon/\partial \mu$ along the phase boundary, we obtain the critical doping
\begin{equation}
  \xc = \sqrt{\Delta U \cdot 2 \kappa (d_\text{m}-d_\text{i})}. \label{eq:xc}
\end{equation}
Similar to the liquid-gas transition, thermodynamics forbids charge densities
lying in the range $0 < |x| < \xc$. The system will phase separate if doped to
lie within this regime~\cite{Galanakis2011}.

We note that the filling-controlled transition is not doping in the
conventional sense, where the insulator is connected to a metal formed by
shifting $\mu$ into the bands lying adjacent to the spectral gap. Indeed the
smallness of $\Delta\mu$ for small $\Delta U$ implied by Eq.~\eqref{eq:1}
dictates that the first order transition occurs without the closing of the
single-particle gap, when $\Delta U$ is small.  Rather, the Mott insulator
transitions to a disconnected, lower-energy, metallic state~\cite{Balzer2009}.


\begin{figure}
  \includegraphics[width=0.6\columnwidth]{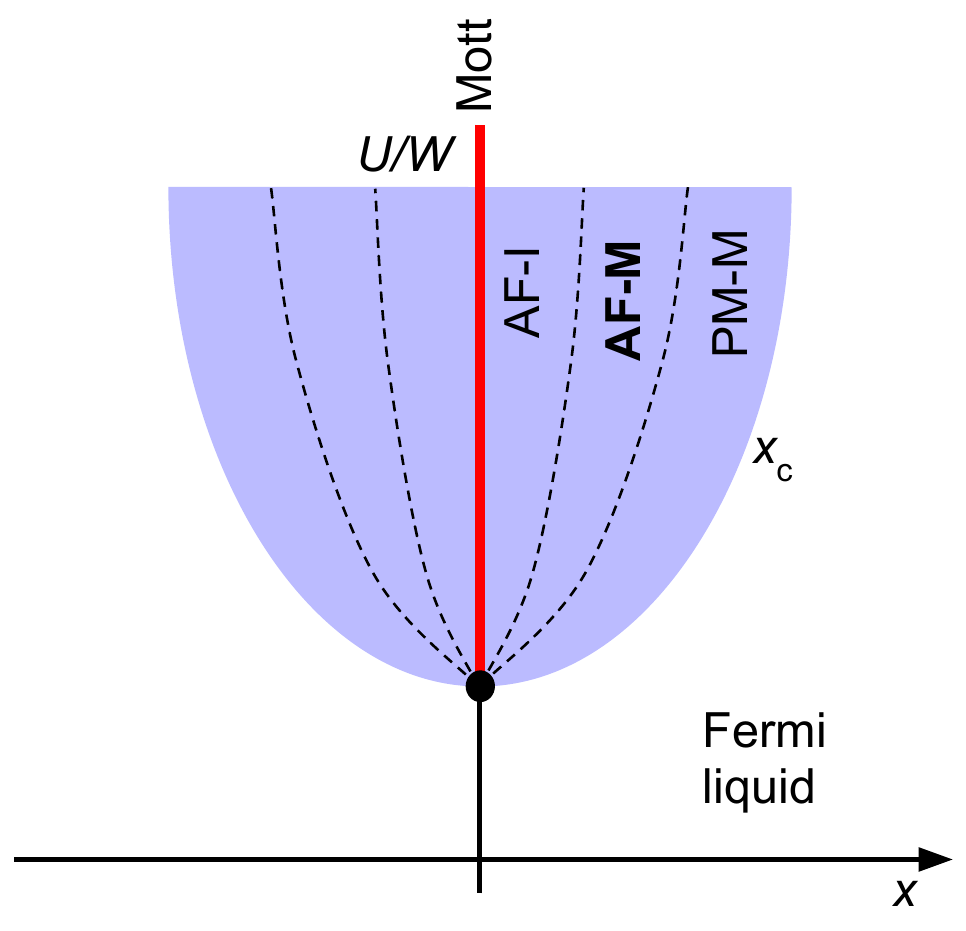}
  \caption{Generic Mott phase diagram for a 3D system plotted in the $U$-vs-$x$
    plane. Beginning at the pure Mott insulating state at zero doping $x = 0$,
    we progress through three phase-separated states (shaded) to arrive at a
    uniform Fermi liquid. The three phase-separated states have distinct
    magnetic (AF or PM) and transport (M or I) signatures. Since the
    percolation threshold $\phi_\text{c}^\text{3D} \sim 1/3$, we expect an
    intermediate phase (AF-M, bolded text) where metallic conductivity coexists
    with magnetic order. This intermediate phase is absent in 2D since
    $\phi_\text{c}^\text{2D} \sim 1/2$, so the metal and insulator are never
    simultaneously percolated.}
  \label{fig:U-vs-n}
\end{figure}

\emph{Phase separation} -- Thermodynamics forbids charge densities in the range
$0 < |x| < \xc$, causing the system to phase separate into insulating regions
with $x = 0$ and metallic regions with $x = \xc$ (shaded region in
Fig.~\ref{fig:U-vs-n}). The surface energy $E_\text{surface} \sim \sigma
L^{d-1}$, where $\sigma > 0$ is the surface tension and $L$ is the
characteristic size of a metallic region, favors forming a single large puddle.
However, the long-ranged part of the Coulomb interaction $E_\text{coul} \sim
\xc^2 L^{2d-1}$ penalizes macroscopic charge imbalances. Balancing the two
gives domains of typical size $L \sim (\sigma/\xc^2)^{1/d}$. The actual spatial
patterns formed depend on system-specific details such as dimensionality,
anisotropy, and elastic forces~\cite{Ortix2009}.

Conducting transport does not coincide with the disappearance of phase
separation at $\xc$ and the formation of the homogeneous metallic state, but
rather when the volume fraction $x/\xc \sim \phi$ of the metallic puddles
reaches the percolation limit, roughly $\phi_\text{c} \sim 1/3$ in three
dimensions~\cite{Essam1980}. Depending on the spatial patterns favored, we may
expect anisotropic transport. Additionally, we predict an intermediate
conducting magnetic state (AF-M in Fig.~\ref{fig:U-vs-n}) since long-range
order persists as long as the insulating regions percolate, up to doping $x/\xc
\sim 1-\phi_\text{c}$. This intermediate state does not exist in two dimensions
since $\phi_\text{c} \sim 1/2$, implying the metallic and insulating states
never simultaneously percolate.

\begin{figure}
  \includegraphics[width=\columnwidth]{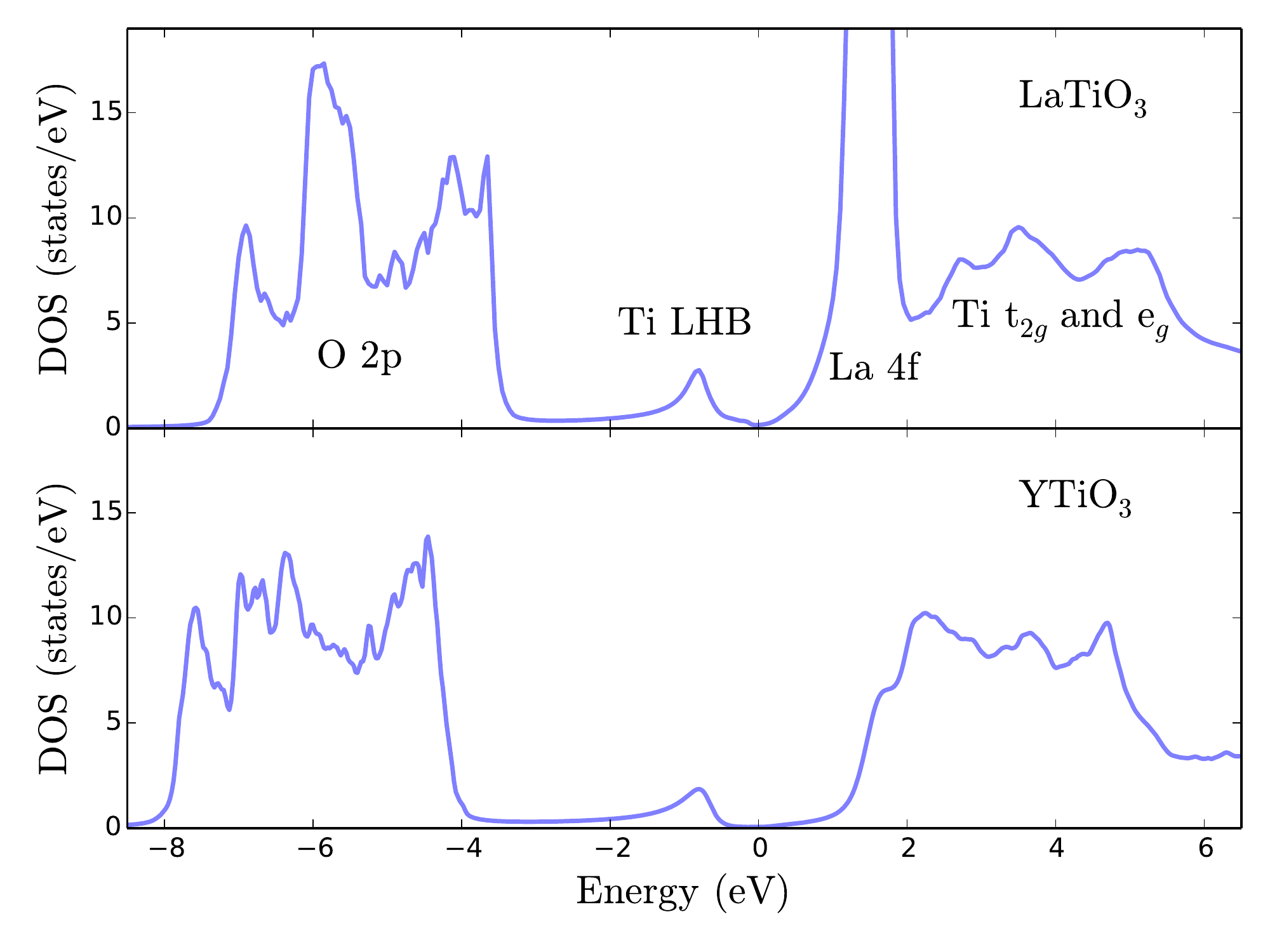}
  \caption{Density of states for end members LaTiO$_3$ and YTiO$_3$ of the
    $R$TiO$_3$ series computed using DFT+DMFT. The reduction of bandwidth in
    YTiO$_3$ enhances the relative strength of correlations and produces a
    larger spectral gap. We emphasize that a single set of Coulomb parameters
    were used for both simulations, and the differences are driven purely by
    chemistry.}
  \label{fig:dos}
\end{figure}

\emph{Ab initio modeling} -- The rare earth titanates $R$TiO$_3$ are an ideal
system to investigate the Mott transition~\cite{Greedan1985,
  Mochizuki2004}. Varying the ionic radius of the rare earth $R$ tunes the
correlation strength, while rare earth vacancies~\cite{Sefat2006} or Ca
substitution~\cite{Katsufuji1997} tunes the Ti valence from $d^1$ to $d^0$. The
interplay between structure, transport and magnetism are
well-characterized. Critical dopings, determined via transport, range from 0.05
in LaTiO$_3$ to 0.35 in YTiO$_3$, and the predicted intermediate metallic
antiferromagnetic state has been observed~\cite{Katsufuji1997}, although the
claim is not without controversy~\cite{Sefat2006a}. Careful bulk measurements
suggest signatures of phase separation~\cite{Hays1999, Zhou2005}. However,
these prior studies suffer from chemical disorder due to the divalent
substitution used to obtain filling control, so recent synthesis of
high-quality electrostatically-doped heterostructures opens the possibility of
filling-control without cation disorder~\cite{Moetakef2011}.

To apply our theory to the titanates, we perform electronic structure
calculations using the combination of density functional theory and dynamical
mean-field theory~\cite{Kotliar2006} with the implementation described in
Ref.~\onlinecite{Haule2010}. We used $U = 9.0$~eV and $J = 0.8$~eV for the
strength of the Coulomb repulsion on the Ti $t_{2g}$ orbitals, and $E_\text{dc}
= U(n_d-1/2) - J(n_d-1)/2$ with $n_d = 1.0$ as the standard double-counting
energy. The empty $e_g$ orbitals do not require correlations for their correct
description. We include all the valence states, notably the oxygen $2p$ states
in the hybridization window. We use $T = 100$~K, well below the Mott transition
temperature at half-filling. The value for $U$ was determined by requiring the
calculated gap of the end-member LaTiO$_3$ to match the
experimentally-determined value, reported to be in the range 20~meV to
0.2~eV~\cite{Lunkenheimer2003}. Once fixed, these parameters were used to for
the entire $R$TiO$_3$ family.  To capture correlations in the $4f$ shells of
the compounds with partially-filled rare earth ions, we applied the atomic
self-energy
\begin{equation}
  \Sigma_f(i\omega_n) = \Sigma_0 + \frac{U_f^2 p (1-p)}{i\omega_n + \mu - U_f(p-1/2)},
\end{equation}
with the static shift $\Sigma_0 = -U_f(p-1/2) - \epsilon_f$. Here, $U_f =
10$~eV is the Hartree term on the $f$-shell, $\epsilon_f$ is the center of mass
of the $f$ density of states, and $p$ is the filling fraction (e.g. $3/14$ for
NdTiO$_3$). Since the chemical potential is the independent variable in the
scans needed to compute the $n$ vs. $\mu$ curves, we do not update the
charge-density, as this would have required self-consistent adjustment of the
nuclear charges. To obtain spectral quantities, we analytically continued the
$3d$ self-energy $\Sigma$ onto the real axis by applying the maximum entropy
method to the effective Green's function $G = 1/(i\omega_n - E -
\Sigma(i\omega_n))$.

Shown in Fig.~\ref{fig:dos} is the density of states for the end-compounds
LaTiO$_3$ and YTiO$_3$. The contraction of the cation ionic radii from La to Y
enhances the octahedral distortions, reducing the bandwidth of YTiO$_3$
relative to LaTiO$_3$ (observed within DFT). The reduction places the YTiO$_3$
deeper inside the Mott insulating state, which is reflected in the increased
spectral gap of nearly 2~eV. The salient features---the location of the lower
Hubbard band and oxygen 2$p$ binding energies---agree well with
photoemission~\cite{Fujimori1992, Fujimori1992a}.

\begin{figure}
  \includegraphics[width=\columnwidth]{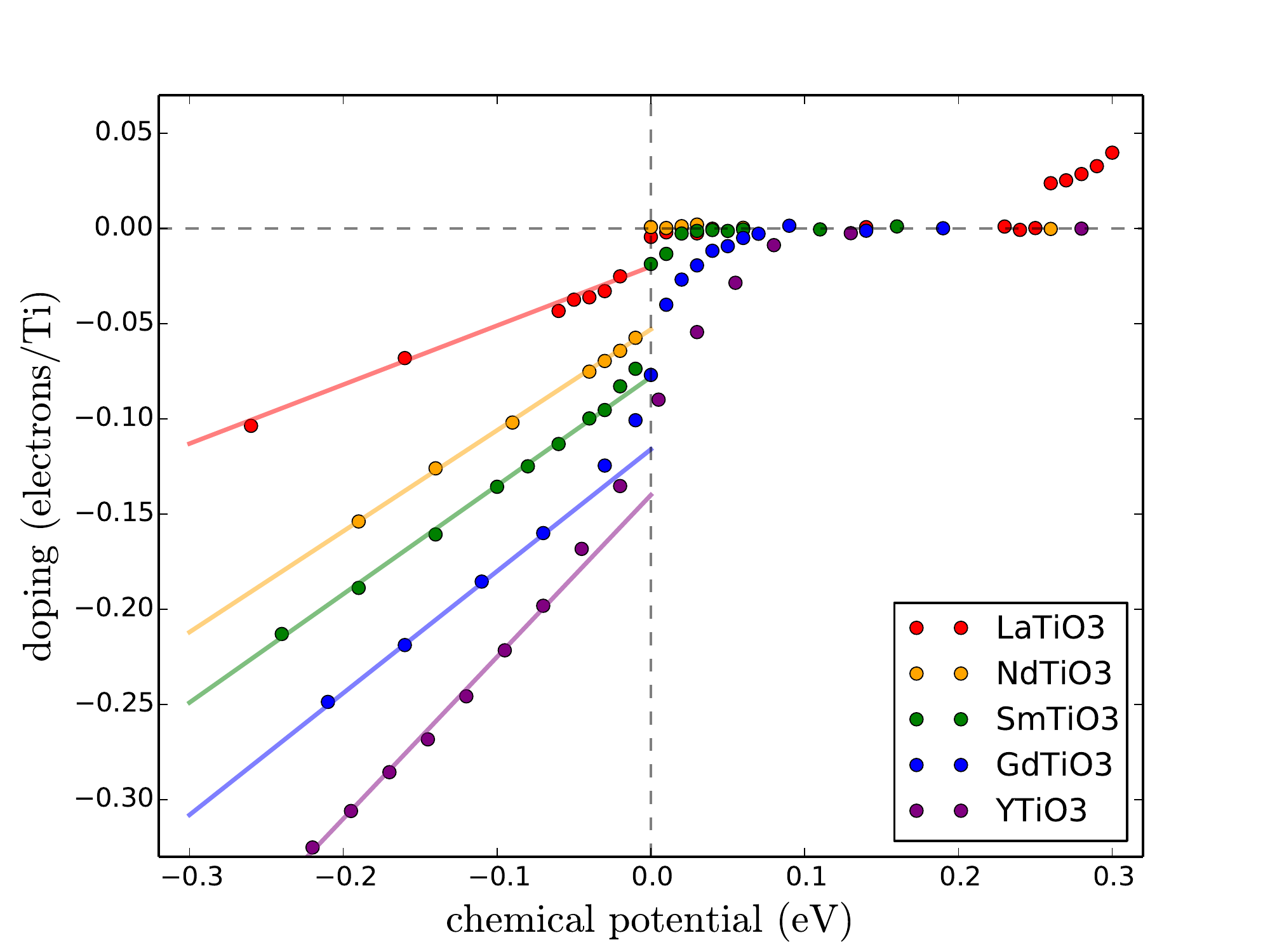}
  \caption{Doping as a function of chemical potential near the hole-doped Mott
    transtion, computed with DFT+DMFT for representative members of the
    $R$TiO$_3$ family. The size of the density discontinuity (the critical
    doping $x_\text{c}$) increases as we progress away from the the largest
    rare earth La. The lines are guides to the eye. The electron-doped
    transition can be seen for LaTiO3 in the upper right.}
  \label{fig:n-vs-mu}
\end{figure}

We explicitly determine the critical doping $\xc$ of the titanates by
monitoring the charge density as we lower the chemical potential to hole-dope
the Mott insulator (Fig.~\ref{fig:n-vs-mu}). The critical doping, as given by
the discontinuity between the insulator and Fermi liquid, increases
monotonically from $\sim 2\%$ for La to $\sim 15\%$ for Y, corroborating our
expectation that correlations increase $\xc$.  We note that the small
contribution to the compressibility due to the partially-filled $4f$ shells for
the intermediate rare earths has been subtracted out to give a flat $n$
vs. $\mu$ curve in the Mott insulating regime. We do not observe a jump in
GdTiO$_3$ and YTiO$_3$ because the Mott critical endpoint drops below the
simulation temperature of $T = 100$~K, as observed
experimentally~\cite{Katsufuji1997}, so we roughly extract $\xc$ by pinpointing
the location of steepest slope in the $n$ vs. $\mu$ curve. The critical dopings
are smaller than experiment a factor of 2, which we attribute to the effect the
strong chemical disorder required for doping, as well as polarons, which is
known to drive the finite-$T$ Mott transition more strongly
first-order~\cite{Capone2004}.


As a consistency check, we also determine $\xc$ for representative compounds
using Eq.~\ref{eq:xc}, which is valid near bandwidth-controlled transition
point.  First, we determined the critical Coulomb strengths $U_\text{c}$ for
the bandwidth-controlled transition, which decrease from LaTiO$_3$ to YTiO3$_3$
as expected. The charge compressibility was obtained by scanning $n$ vs. $\mu$
at $U_\text{c}$. To obtain the ``double-occupancy'' of the metallic and
insulating solutions, we note that in multiband models, the Coulomb $U$ couples
to the generalization of the on-site double-occupancies---the Hartree component
of the potential energy---$N_i(N_i-1)/2$ where $N_i$ runs from 0 to 10 within
the 3$d$ manifold. The extracted parameters are shown in
Table.~\ref{tbl:params}. Again, $\xc$ increases as we progress from the least-
to the most-correlated compounds and roughly agree with those from the $n$
vs. $\mu$ curves, even for YTiO$_3$ which is quite far from the
bandwidth-controlled transition.


\begin{table}
  \centering
  \begin{tabular}{r|ccc|ccc}
    Compound & $\kappa$ ($e$/eV) & $d_\text{m}$ & $d_\text{i}$ & $U_\text{c}$ (eV) & $\xc$ \\
    \hline
    LaTiO$_3$ & 0.20 & 0.15  & 0.13  & 8.8 &   4\% \\
    SmTiO$_3$ & 0.22 & 0.22  & 0.19  & 6.0 &  20\% \\
    YTiO$_3$  & 0.28 & 0.23  & 0.20  & 4.7 &  27\% \\
  \end{tabular}
  \caption{For representative titanates, we tabulate the electronic
    compressibility per Ti atom $\kappa = \partial n/\partial \mu$, Hartree
    component of the potential energy $d_\text{m,i} = \avg{N(N-1)/2}$ in the
    metallic and insulating states, and the critical Coulomb strengths
    $U_\text{c}$. Using $\Delta U = U - U_\text{c}$ where $U = 9$~eV in our
    calculations, and Eq.~\ref{eq:xc}, we compute the critical doping $\xc$.}
  \label{tbl:params}
\end{table}

\emph{Summary} -- We have outlined a theory for the first-order
filling-controlled Mott transition, which predicts intrinsic electronic phase
separation when a Mott insulator is doped away from half-filling, and
demonstrated explicitly how to calculate the critical doping $\xc$ in
electronic structure calculations. The thermodynamic signatures of this
pervasive phase-separation has been observed in many other correlated
systems~\cite{Imada1998}, as well as directly using near-field optics on
VO$_2$~\cite{Qazilbash2007} and STM in the cuprates~\cite{Kohsaka2012}. The key
tasks to enhance the quantitative agreement between theory and experiment
involve (a) including disorder and polarons into theoretical calculations, and
(b) designing cleaner experimental systems where chemical disorder can be
reduced, e.g. through modulation-doped samples or oxide heterostructures. The
accessibility of thin films to spatially resolved probes (STM,
spatially-resolved optics) is especially advantageous as they would allow
direct visualization of the phase separated region.

C.Y. was supported by the Army Research Office MURI grant W911-NF-09-1-0398.
L.B.  was supported by the MRSEC Program of the National Science Foundation
under Award No. DMR 1121053. We benefitted from facilities of the KITP, funded
by NSF grant PHY-11-25915. We also acknowledge support from the Center for
Scientific Computing at the CNSI and MRL: an NSF MRSEC (DMR-1121053) and NSF
CNS-0960316.

\bibliography{doped-mott-insulators}

\end{document}